\documentclass[twocolumn, aps,prl, showpacs,preprintnumbers,bibnotes]{revtex4-1}
\pdfoutput=1
\usepackage[T1]{fontenc}

\usepackage{graphicx}
\usepackage{bm}
\usepackage{color}
\usepackage[T1]{fontenc}
\usepackage{amsmath}
\usepackage{upgreek}
\usepackage{nicefrac}

\newcommand\unit[2]{\ensuremath{#1~\mathrm{{#2}}}}
\newcommand\Ket[1]{\ensuremath{|{#1}\rangle}}

\newcommand\Li{\ensuremath{{^6}\mathrm{Li}}}

\begin{document}

\title{Site-resolved Imaging of Fermionic \Li{} in an Optical Lattice}
\author{Maxwell F. Parsons}
\author{Florian Huber}
\author{Anton Mazurenko}
\author{Christie S. Chiu}
\author{Widagdo Setiawan}
\author{Katherine Wooley-Brown}
\author{Sebastian Blatt}
\author{Markus Greiner}
\email{greiner@physics.harvard.edu}
\affiliation{
  Department of Physics, Harvard University,
  Cambridge, Massachusetts, 02138, USA}

\date{\today}

\begin{abstract}
We demonstrate site-resolved imaging of individual fermionic \Li{} atoms in a 2D optical lattice.  To preserve the density distribution during fluorescence imaging, we simultaneously cool the atoms with 3D Raman sideband cooling.  This laser cooling technique, demonstrated here for the first time for \Li{} atoms, also provides a pathway to rapid low-entropy filling of an optical lattice.  We are able to determine the occupation of individual lattice sites with a fidelity \textgreater 95\%, enabling direct, local measurement of particle correlations in Fermi lattice systems.  This ability will be instrumental for creating and investigating low-temperature phases of the Fermi-Hubbard model, including antiferromagnets and d-wave superfluidity.

\end{abstract}

\pacs{03.75.Ss, 07.60.Pb, 37.10.De}
\maketitle

Particle correlations reveal the underlying order of an interacting quantum many-body system.  Strong correlations give rise to rich quantum many-body phenomena such as high-temperature superconductivity and colossal magneto-resistance \cite{Dagotto2005}.  One approach toward studying correlated many-body systems uses ultracold atoms to implement a well-understood and tunable realization of a particular model, and to use the behavior of the clean atomic system as a benchmark for theory \cite{Bloch2012}.  This ``synthetic matter'' approach is especially fruitful for strongly-correlated fermionic systems, where, for even the simplest models, the sign problem of the Quantum Monte Carlo method precludes accurate computations of thermodynamic observables \cite{Troyer2005}.  In addition to theoretical simplicity and tunability, ultracold atomic systems can be designed to have interparticle spacings of order the wavelength of visible light.  By placing a quantum gas under an optical microscope we can therefore directly observe and manipulate quantum correlations at their smallest length scale.  Such a quantum gas microscope has been realized for bosonic $^{87}$Rb \cite{Bakr2009,Sherson2010} and $^{174}$Yb \cite{Miranda2015} atoms.   In bosonic systems, site-resolved imaging has been used to study the quantum phase transition from a superfluid to a Mott insulator \cite{Bakr2010, Sherson2010, Endres2012} and from a paramagnet to an antiferromagnet \cite{Simon2011}.  Single-site resolution also enables the extraction of non-local order parameters such as string order \cite{Endres2011} and allows studies of strongly-correlated dynamics in optical lattices \cite{Preiss, Fukuhara2013, Cheneau2012}.  Until very recently \cite{Huber2014, Haller2015, Cheuk}, however, site-resolved imaging had not been demonstrated for fermionic atoms.  In Fermi-Hubbard systems, cold atom experiments without single-site resolution have observed Mott insulators \cite{Jordens2008, Schneider2008} and antiferromagnetic correlations \cite{Greif2013a, Hart2014}.  In these experiments, understanding of the prepared many-body state is limited by lack of direct access to the many-body wave function and the inability to locally measure correlations.  The extension of quantum gas microscopy to fermions will provide novel probes for Fermi lattice systems, such as site-resolved spin correlation functions and local entropy measurement.

\begin{figure}[t]
\centering
\includegraphics{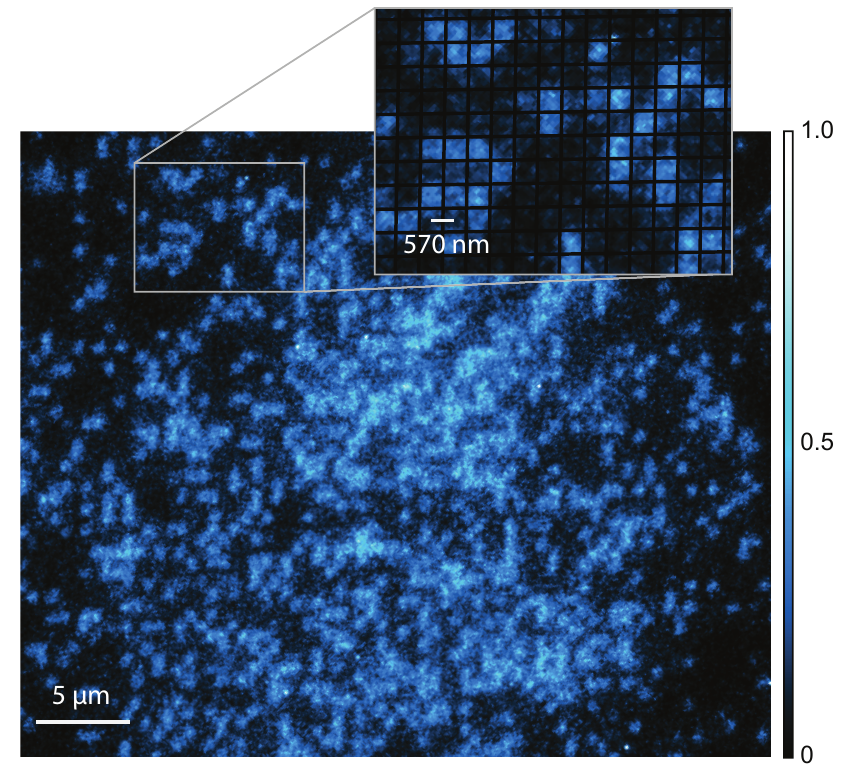}
\caption{(color online). Fluorescence image of atoms in a single layer of a cubic lattice obtained using Raman sideband cooling.  The filling fraction in the center of the cloud is 40\%. We collect approximately 750 photons per atom during a \unit{1.9}{s} exposure.  The colorbar is in arbitrary units.}
\label{fig:dense-lattice}
\end{figure}

Here, we demonstrate site-resolved imaging of fermionic \Li{} in a 2D optical lattice with high fidelity  [see Fig.~\ref{fig:dense-lattice}]. \Li{} is an especially suitable species for many-body experiments with ultracold atoms because its light mass leads to fast thermalization and dynamics, and its broad magnetic Feshbach resonances \cite{Chin2010} allow precise control of atomic interactions.  The natural energy scale for particles of mass $m$, in an optical lattice with spacing $a$, is the recoil energy,  $\mathrm{E_{r}} = h^{2}/8 a^{2} m$, where $h$ is Planck's constant.  For many-body physics, working with a light atom gives an advantage because the recoil energy scales inversely with the mass.  Experiments studying antiferromagnetic correlations with $^{40}$K \cite{Greif2013a} have been limited by heating, owing to the intrinsic slow dynamics of cold atoms.  The natural timescale for \Li{} is 7 times faster than for $^{40}$K in a system with identical lattice geometry.  For microscopy, however, the light mass creates a challenge because the recoil energy due to photon scattering also scales inversely with the atomic mass, requiring very large trap depths for imaging.  We overcome this challenge by implementing 3D Raman sideband cooling \cite{Monroe1995,Hamann1998,Kerman2000,Han2000,Thompson2013,Kaufman2012,Lester2014,Patil2014} for \Li{} atoms in a \unit{2.4}{mK} deep optical lattice.

\begin{figure}[t]
\centering
\includegraphics{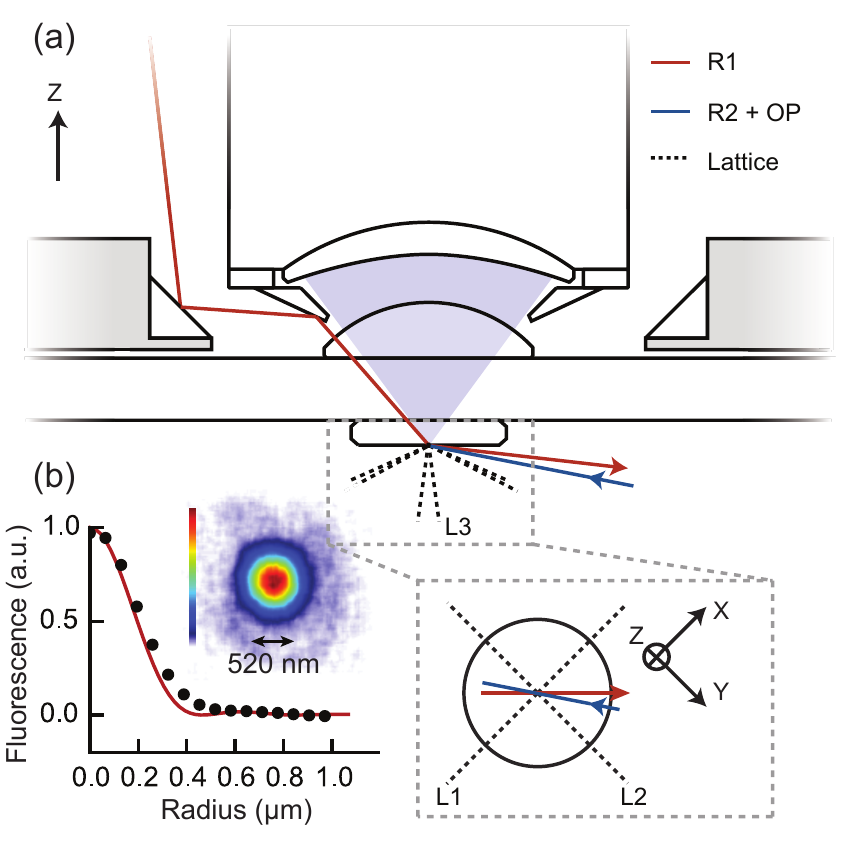}
\caption{(color online). A schematic of the microscope.  R1 and R2 denote our Raman beams, and OP the optical pumping light which co-propagates with R2.  L1 and L2 are additionally retroreflected out of the schematic to create a 3D lattice as described in Ref. \cite{Blatt2015}.  L3 forms a lattice along $\hat{z}$, providing additional confinement during imaging.  L1, L2, and L3 have waists of \unit{80}{\mu m}, \unit{80}{\mu m}, and \unit{40}{\mu m}, respectively.  The measured point spread function, obtained by superimposing and averaging isolated atoms, is shown in panel (b).  The black markers are an azimuthal average of the measured point spread function (PSF).  The red curve is the expected diffraction-limited Airy disk for an NA of 0.87.  The inset is an image of the PSF.  A gaussian fit to the PSF yields a full width at half maximum of \unit{520}{nm}, compared to our lattice spacing of \unit{569}{nm}.}
\label{fig:schematics}
\end{figure}

Atoms are trapped in a vacuum glass cell, \unit{9.9}{\mu m} beneath the surface of a superpolished substrate, in the object plane of a 0.87 numerical aperture (NA) imaging system.  Our imaging system combines a long working distance microscope objective (Optem 20X, NA$=$0.6) with a hemispherical lens to enhance the NA.  We compensate spherical aberration with a phase plate in the imaging system.  We image the atomic fluorescence onto the photocathode of a gateable intensified CCD camera (Andor iStar 334T) with a magnification of 170.  We achieve diffraction-limited resolution, shown in Fig.~\ref{fig:schematics}(b).  The full width at half maximum from a Gaussian fit to the measured point spread function is \unit{520}{nm} compared to a lattice spacing of \unit{569}{nm}.

Atoms in an equal mixture of \Ket{F=\nicefrac{1}{2}, m_{\mathrm{F}}=\pm \nicefrac{1}{2}} in the electronic ground state are loaded from a single layer of a 1D ``accordion lattice'' with tunable spacing into a 2D optical lattice (see Ref.  \cite{Blatt2015}).  Lattice beams L1 and L2 [see Fig.~\ref{fig:schematics}(a)] form radial lattices along $\hat{x}$ and $\hat{y}$, respectively, with 569 nm spacing. L1 and L2 also each form an axial lattice along $\hat{z}$ with \unit{1.48}{\mu m} spacing.  During the initial lattice loading, L1 and L2 are each ramped up in \unit{100}{ms} to give radial lattice depths of 30 $\mathrm{E_{r, rad}}$, where tunneling is suppressed.  For imaging, we introduce an additional lattice along $\hat{z}$,  with \unit{534}{nm} spacing, formed by L3.  All lattices are derived from \unit{1064}{nm} light.  Just before imaging, L1, L2, and L3 are ramped in 100 ms to give nearly-degenerate on-site trap frequencies of $(\omega_{x}, \omega_{y},\omega_{z}) = 2\pi \times (\unit{1.25}{MHz}, \unit{1.25}{MHz}, \unit{1.47}{MHz})$, calibrated using lattice modulation spectroscopy.

\begin{figure}[b]
\centering
\includegraphics{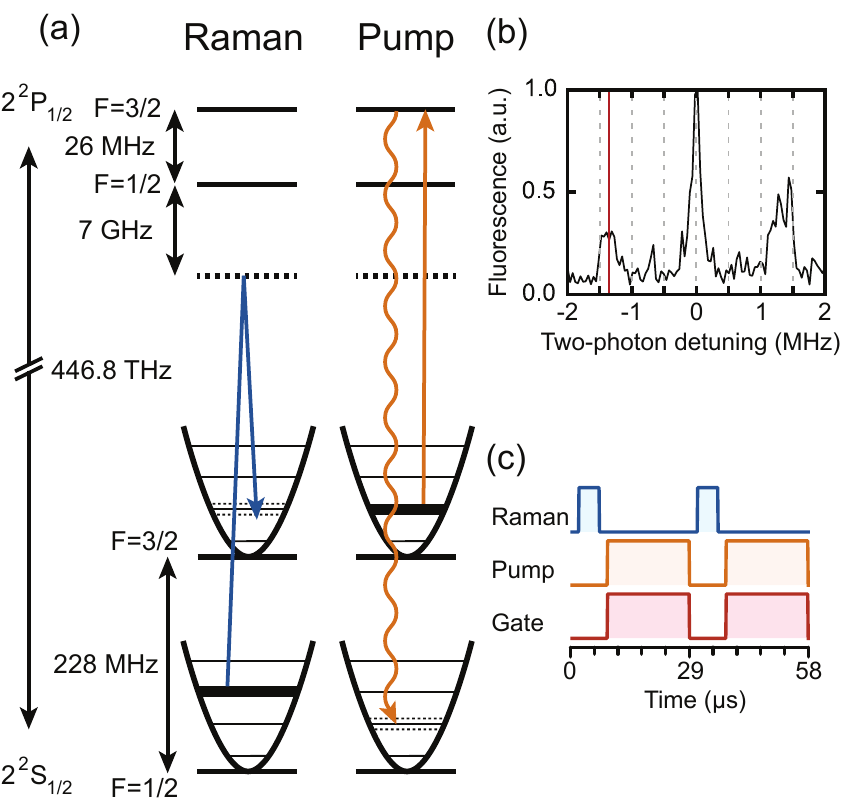}\\
\caption{(color online). Pulsed Raman sideband imaging. A Raman transition drives atoms into the \Ket{2^2\mathrm{S}_{1/2}(F=\nicefrac{3}{2})} hyperfine manifold, removing one vibrational excitation.  Atoms are then optically pumped back into the \Ket{2^2\mathrm{S}_{1/2}(F=\nicefrac{1}{2})} manifold while simultaneously switching on the intensifier of an intensified CCD camera to collect the photons scattered during pumping.  A spectrum, taken by driving a Raman transition with a $\unit{200}{\mu s}$ long pulse and then imaging the $\Ket{2^2\mathrm{S}_{1/2}(F=\nicefrac{1}{2}, m_{\mathrm{F}} = \nicefrac{-1}{2})}$ state is shown in panel (b), with the red line denoting the two-photon detuning during imaging.  The timing of two imaging pulses is shown in panel (c).}
\label{fig:raman}
\end{figure}

To keep the atoms pinned to their lattice sites during fluorescence imaging we must simultaneously cool them.  Previous quantum gas microscopes have used a polarization gradient cooling (PGC) scheme for imaging $^{87}$Rb \cite{Bakr2009, Sherson2010}.  PGC is not suitable for sub-Doppler cooling of \Li{} due to the unresolved hyperfine splitting in the excited state \cite{Duarte2011}.  Sisyphus cooling has been demonstrated for \Li{} in free space \cite{Hamilton2014} and gray-molasses cooling has been demonstrated for \Li{} both in free space and in an optical dipole trap \cite{Burchianti2014}.  These cooling techniques, however, have not yet been extended to the tightly-confined regime of optical lattices with \Li{}.  We use Raman sideband cooling because it does not rely on resolved hyperfine structure and has been demonstrated to cool a variety of atomic species to the motional ground state in optical lattices \cite{Hamann1998, Kerman2000}, optical tweezers \cite{Thompson2013, Kaufman2012}, and ion traps \cite{Monroe1995}, as well as to image $^{87}$Rb atoms in optical tweezers \cite{Lester2014} and optical lattices \cite{Patil2014}.

To image the atoms we collect the photons scattered during optical pumping in the pulsed Raman sideband cooling scheme shown in Fig.~\ref{fig:raman}.  The imaging is performed at a magnetic field of $<\unit{20}{mG}$.  First, a Raman transition drives the atoms into \Ket{2^2\mathrm{S}_{1/2}(F=\nicefrac{3}{2})}, removing one vibrational excitation. The Rabi frequency for a Raman cooling transition on the lowest motional sideband for lattice axis $\nu$ is given by $\eta_{\nu} \Omega_{\mathrm{c}}$, where $\eta_{\nu} = \delta k_{\nu} x_{\nu} = (0.47, 0.47, 0.15)$.  Here, $\eta_{\nu}$ is the Lamb-Dicke parameter for the Raman transition, $\delta k_{\nu}$ is the projection of the difference in the Raman beam wave vectors along the lattice axes, $x_{\nu}$ is the harmonic oscillator length, and $\Omega_{\mathrm{c}} = 2 \pi \times \unit{160}{kHz}$ is the two-photon Rabi frequency on the carrier.  The Raman beams have linear polarization to avoid effective magnetic fields. During the Raman pulse, the camera intensifier is gated off to suppress background from the Raman light.  After a \unit{5}{\mu s} Raman pulse, the atoms are pumped with resonant light through  \Ket{2^2\mathrm{P}_{1/2}(F=\nicefrac{3}{2})} back into the  \Ket{2^2\mathrm{S}_{1/2}(F=\nicefrac{1}{2})} dark state at a rate of $\sim$\unit{1.5 \times 10^5}{s^{-1}} for \unit{20}{\mu s}, completing one imaging pulse.  The camera intensifier is gated on during the optical pumping step to collect the scattered photons and form an image.  To obtain one image with $\sim$750 photons collected per atom, we apply $6.4\times 10^{4}$ imaging pulses over \unit{1.9}{s}.

\begin{figure}[b]
\centering
\includegraphics{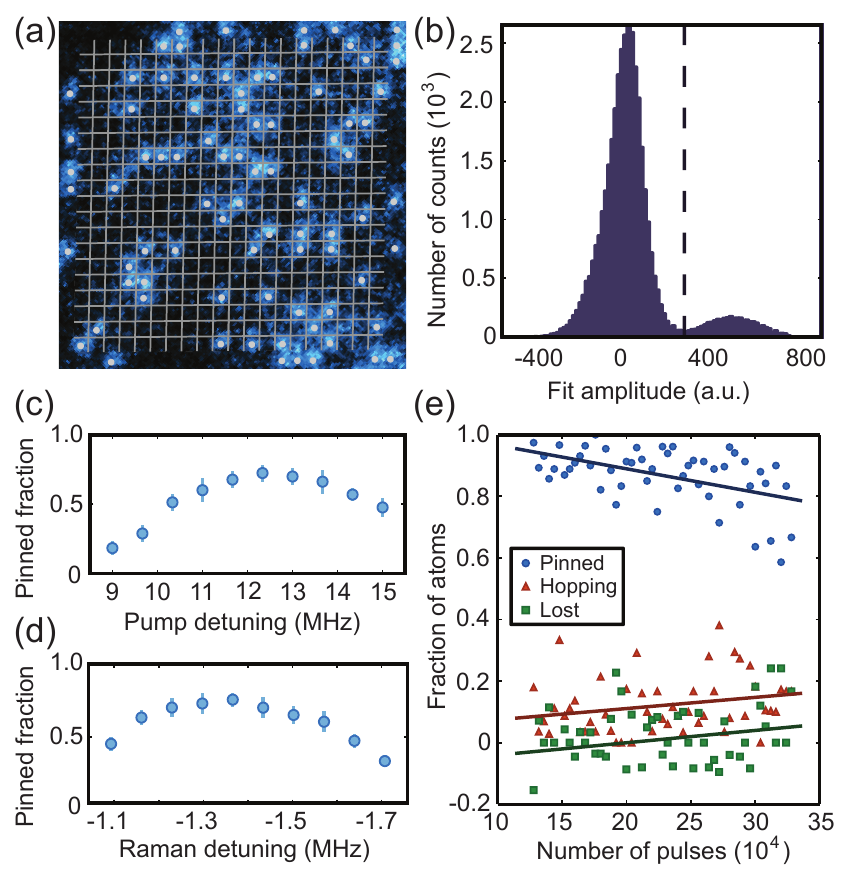}\\
\caption{(color online). Site-resolved imaging with high fidelity.  Images obtained by applying $6.4\times 10^{4}$ Raman imaging pulses are fit to a lattice of the measured point spread function, with the amplitudes for each lattice site as fit parameters (a).  Whether a site is occupied is determined by applying a threshold to the fitted amplitudes, with occupied sites denoted by white dots.  An averaged histogram of fitted amplitudes is shown in panel (b).  We optimize the imaging by taking two subsequent frames, with additional imaging pulses in between, and looking at the fraction of atoms that remain pinned to their sites between the two images (c, d).  By varying the number of pulses between frames and applying a linear fit to the pinned, hopping and lost fractions, we extract loss and hopping rates (e).  As expected, within the statistical uncertainty of the fit, the x-intercepts of the hopping and lost fractions occur at $6.4\times 10^{4}$ pulses (equivalent to one frame).}
\label{fig:fidelity}
\end{figure}

For efficient cooling the system must be in the Lamb-Dicke regime, $\eta_{\mathrm{OP}}=k_{\mathrm{OP}} x_{\nu} \ll 1$, where the optical pumping process preserves the vibrational state with high probability.  Here, $\eta_{\mathrm{OP}} \approx 0.31$ is the Lamb-Dicke parameter for the pumping process, and $k_{\mathrm{OP}}$ is the magnitude of the wavevector for the pump light.  Achieving the Lamb-Dicke regime for \Li{} requires MHz-level trap frequencies, which are atypically large for neutral atom experiments \cite{Blatt2015}.  The small lattice beam waists in the experiment cause inhomogeneity of the trap frequency over the sample size.  The lattice along $\hat{z}$ has the largest inhomogeneity, with the trap frequency varying by \unit{120}{kHz} over a radius of 30 lattice sites.  We have found that the imaging works optimally for Raman pulse durations of \unit{5}{\mu s}, where Fourier-broadening exceeds the inhomogeneity in trap frequency.  Additionally, we find a strong dependence of the imaging fidelity on the detuning of the optical pumping light [see Fig.~\ref{fig:fidelity}(c)].  The optimal pump detuning is in agreement with the expected shift of the pump resonance due to the AC stark shift in the lattice, based on the polarizabilities calculated in \cite{Safronova2013}.

	We reconstruct the atom distribution in the lattice by fitting images to a lattice of the measured point spread function (PSF) [see Fig.~\ref{fig:fidelity}(a)].  The PSF of the imaging system and the lattice geometry are determined once from images of a sparsely-filled lattice and used for fitting subsequent images.  Each image is divided into 10$\times$10-site subregions, and each subregion is fitted with PSF amplitudes for each site, a uniform background, and a global 2D coordinate offset as fit parameters.  A threshold is then applied to the fitted amplitudes to determine which sites were occupied.  A histogram of the fitted amplitudes [see Fig.~\ref{fig:fidelity}(b)] shows a bimodal distribution with the peaks corresponding to unoccupied and occupied sites.  We do not observe peaks corresponding to more than one atom per site because pairs of atoms are ejected during imaging due to light-assisted collisions \cite{Bakr2009}.  Both the motion of atoms between lattice sites during imaging and the quality of the image fit contribute to the imaging fidelity.  By simulating images---taking into account photon shot noise, camera noise, image background, and the measured variance in atom fluorescence---we evaluate the accuracy of the density reconstrunction algorithm alone, isolated from the effects of atomic motion.  The accuracy is determined by comparing the known density distribution in simulated images with the results from applying our fitting algorithm to the same images.  For a lattice with 20\% of the sites occupied, we find that the algorithm correctly identifies occupied sites $(98.7 \pm 0.5)$\% of the time and correctly identifies unoccupied sites $(99.7 \pm 0.2)$\% of the time.

To study atom hopping and loss due to the imaging, we take two images with $6.4 \times 10^{4}$ Raman imaging pulses each and apply a varying number of Raman imaging pulses applied in between them.  By comparing the reconstructed atom distribution of the two frames, we determine the fraction of atoms that stay pinned to their sites, hop between sites, and are lost from the image ($f_{\mathrm{p}}$, $f_{\mathrm{h}}$, and $f_{\mathrm{l}}$).  Loss can be caused by atoms leaving the region of analysis, hopping along $\hat{z}$, or leaving the trap.  The Raman imaging parameters are optimized on the pinned fraction [see Fig.~\ref{fig:fidelity}(c, d)].  Fig.~\ref{fig:fidelity}(e) shows $f_{\mathrm{p}}$, $f_{\mathrm{h}}$, and $f_{\mathrm{l}}$ versus the total number of imaging pulses for optimized imaging parameters.  By applying a linear fit to these data, we determine rates that we use to get the expected pinned, hopping, and lost fractions for a single image with $6.4 \times 10^{4}$ pulses.  For a single image we have $f_{\mathrm{p}}  = (95.1 \pm 1.2)$\%, $f_{\mathrm{h}} = (2.3 \pm 1.3)$\%, and  $f_{\mathrm{l}} = (2.6 \pm 1.7)$\% in a 20$\times$20-site analysis region. A negative lost fraction corresponds to atoms entering the region of analysis.  In a lattice with unity filling, each hopping event will cause the loss of two atoms due to light-assisted collisions on doubly occupied sites.  Atoms have uniform probability of hopping at any time during the imaging process.  From the histogram, we see that an atom which hops in the last half of the imaging sequence will still be counted by the density reconstruction algorithm.  We estimate the probability of accurately determining the occupation of a lattice site to be \textgreater 95\%.

In conclusion, we have demonstrated site-resolved detection of fermionic \Li{} in a 2D optical lattice with high fidelity using 3D Raman sideband cooling.  The microscope will provide exquisite control of optical potentials, enabling single-atom addressability and perhaps a route to lower entropy samples \cite{Bernier2009, Lubasch2011}.  The extension of quantum gas microscopy to fermionic systems enables local measurement of particle correlations and will allow new experimental comparisons to the predictions of interacting quantum many-body models.

We would like to thank Eric Tai for assistance with the lattice fitting algorithm, and the Ketterle and Zwierlein groups, Immanuel Bloch, Leslie Czaia, Daniel Greif, Adam Kaufman, Mikhail Lemeshko, Marianna Safronova, Jeff Thompson, Tobias Tiecke, and Vladan Vuleti\'c for helpful discussions.  We acknowledge support from ARO DARPA OLE, AFOSR MURI, ONR DURIP, and NSF.  M.F.P., A.M., and C.S.C. were supported by the NSF GRFP.  S.B. acknowledges support from the Harvard Quantum Optics Center.

\bibliography{singlesite}

\end{document}